\documentclass[aps,prl,twocolumn,superscriptaddress,preprintnumbers,%
               showpacs,nofootinbib]{revtex4}
\newcommand{\PRE}[1]{}       

\usepackage{bm}
\usepackage{epsfig}

\newcommand{\postscript}[2]{\setlength{\epsfxsize}{#2\hsize}
   \centerline{\epsfbox{#1}}}

\newcommand{\amu}{a_{\mu}}
\newcommand{\dmu}{d_{\mu}}

\newcommand{\ecm}{e~\text{cm}}

\newcommand{\eqref}[1]{Eq.~(\ref{#1})}

\newcommand{\bold}[1]{{\text{\normalsize\bm{$#1$}}}}

\begin{document}

\preprint{UCI-TR-2002-29, UFIFT-HEP-02-26, CERN-TH/2002-187}

\title{
\PRE{\vspace*{1.5in}}
The Measurement of the Muon's Anomalous Magnetic Moment Isn't
\PRE{\vspace*{0.3in}}
}

\author{Jonathan L.~Feng}
\affiliation{Department of Physics and Astronomy,
University of California, Irvine, CA 92697, USA
\PRE{\vspace*{.1in}}
}

\author{Konstantin T.~Matchev}
\affiliation{Department of Physics, University of Florida, Gainesville,
FL 32611, USA 
\PRE{\vspace*{.1in}}
}
\affiliation{Theory Division, CERN, CH-1211 Geneva, Switzerland
\PRE{\vspace*{.1in}}
}

\author{Yael Shadmi%
\PRE{\vspace*{.2in}}
}
\thanks{Incumbent of a Technion Management Career Development Chair.}
\affiliation{Department of Physics, Technion, Haifa 32000, Israel
\PRE{\vspace*{.1in}}
}


\begin{abstract}
\PRE{\vspace*{.1in}} 
Recent results announced as measurements of the muon's anomalous
magnetic moment are in fact measurements of the muon's anomalous spin
precession frequency.  This precession frequency receives
contributions from both the muon's anomalous magnetic and electric
dipole moments.  We note that all existing data cannot resolve this
ambiguity, and the current deviation from standard model predictions
may equally well be interpreted as evidence for new physics in the
muon's anomalous magnetic moment, new physics in the muon's electric
dipole moment, or both.
\end{abstract}

\pacs{PACS: 13.40.Em, 14.60.Ef}

\maketitle

Recently the Muon $(g-2)$ Collaboration announced a new measurement of
the muon's anomalous magnetic moment~\cite{Bennett:2002jb}.  More
precisely, however, what has been measured is the muon's anomalous
spin precession frequency.  This receives contributions from both the
muon's anomalous magnetic and electric dipole moments, and we point
out that the reported data and all existing constraints cannot
distinguish between the two.

The recent result is the latest {\em tour de force} from the Muon
$(g-2)$ Experiment~\cite{Carey:dd}.  This experiment measures the
anomalous spin precession frequency of muons circulating in a
perpendicular and uniform magnetic field.  For fermions with
gyromagnetic ratio $g=2$, the cyclotron and spin precession
frequencies are identical.  Measurements of the anomalous spin
precession frequency have therefore been reported as measurements of
the anomalous magnetic dipole moment (MDM) $\amu = (g_{\mu}-2)/2$.

The spin precession frequency is also sensitive to the muon's electric
dipole moment (EDM), however~\cite{Bailey:1979mn,Feng:2001sq}. For a
muon traveling with velocity $\bold{\beta}$ perpendicular to both a
magnetic field $\bold{B}$ and an electric field $\bold{E}$, the
anomalous spin precession vector is
\begin{eqnarray}
\lefteqn{\bold{\omega}_a = -\amu \frac{e}{m_{\mu}} \bold{B}
- \dmu \frac{2c}{\hbar} \bold{\beta} \times \bold{B}} \cr 
&&- \frac{e}{m_{\mu}c} \Bigl(\frac{1}{\gamma^2-1} - \amu\Bigr) 
\bold{\beta} \times \bold{E} 
- \dmu \frac{2}{\hbar} \bold{E} \ ,
\label{omega}
\end{eqnarray}
where $m_{\mu}$ and $\dmu$ are the muon's mass and EDM.  In recent
experiments, the $\bold{\beta} \times \bold{E}$ term of \eqref{omega}
is removed by running at the `magic' $\gamma \approx 29.3$, and the
last term is negligible.  For highly relativistic muons with
$|\bold{\beta}| \approx 1$, then, the anomalous spin precession
frequency is
\begin{equation}
|\bold{\omega}_a| 
\approx |\bold{B}| \biggl[ \Bigl( \frac{e}{m_{\mu}} \Bigr)^2
a_\mu^2 + 
\Bigl(\frac{2c}{\hbar}\Bigr)^2 d_\mu^2 \biggr]^{1/2} \ ,
\label{both}
\end{equation}
and it constrains only a {\em combination} of $\dmu$ and $\amu$.

\begin{figure}[tbp]
\postscript{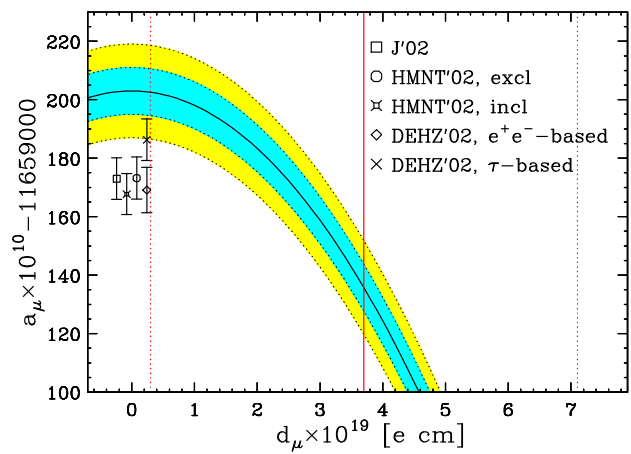}{0.95}
\caption{Regions of the $(\dmu,\amu)$ plane consistent with the
measured $|\bold{\omega}_a|$ at $1\sigma$ (dark) and $2\sigma$
(light).  The most recent SM $\amu$
predictions~\cite{Jegerlehner,Teubner,Davier:2002dy} are also shown,
with horizontal offsets to the SM prediction of $\dmu\approx0$
inserted for clarity.  The vertical lines are the central value and
$\pm 1\sigma$ direct bounds on $\dmu$~\protect\cite{Bailey:1979mn}.
\label{fig:gm2exp}}
\end{figure}

In Fig.~\ref{fig:gm2exp} we show regions of the $(\dmu,\amu)$ plane
that are consistent with the new $|\bold{\omega}_a|$ measurement.
Also shown are the latest standard model (SM)
predictions~\cite{Jegerlehner,Teubner,Davier:2002dy}.  Assuming a
negligible $\dmu$, the measurement shows tentative evidence for new
physics in $\amu$ with uncertain significance, given the spread in
theoretical predictions. However, the $|\bold{\omega}_a|$ result could
just as well be taken as evidence for new physics in $\dmu$.  The best
direct bound on $\dmu$~\cite{Bailey:1979mn} is also shown.  Clearly it
does not resolve this ambiguity; if anything, it favors the EDM
interpretation.  In fact, even taking the lowest SM prediction for
$a_{\mu}$, a striking and unambiguous conclusion is that, barring a
fine-tuned cancelation, the Muon $(g-2)$ Experiment has now set the
most stringent upper bound on the muon's {\em electric} dipole moment
with $|\dmu| < 3.2 \times 10^{-19}~\ecm$.

The MDM/EDM ambiguity may be resolved by appealing to theoretical
prejudice that $\dmu$ is small.  In supersymmetry, for example, the
maximal value of $\amu$ is $\amu^{\text{max}} \sim 10^{-7}$, assuming
only flavor conservation~\cite{Feng:2001tr}. By a phase rotation of
the relevant operator, this implies a maximal EDM of roughly
$\dmu^{\text{max}} \sim (e \hbar / 2 m_{\mu} c) \amu^{\text{max}} \sim
10^{-20}~\ecm$.  This conclusion is far from universal, however.  For
example, large muon EDMs are possible in models where EDMs scale
approximately as $d_f \propto m_f^3$~\cite{Babu:2000cz}.  Given our
current profound ignorance of the origins of electroweak symmetry
breaking, flavor, and CP violation, no definitive statement can be
made.

The effects of $\dmu$ and $\amu$ are, of course,
distinguishable~\cite{Bailey:1979mn}: $\amu$ causes precession around
the magnetic field's axis, but $\dmu$ leads to oscillation of the
muon's spin above and below the plane of motion.  A search for up-down
asymmetry in the current data is in progress~\cite{miller}.  A
dedicated EDM experiment~\cite{Semertzidis:1999kv} would provide a
conclusive resolution by either measuring a non-vanishing $\dmu$ or
constraining the contribution of $\dmu$ to $|\bold{\omega}_a|$ to be
insignificant.

For now, however, the reported data is not a model-independent
measurement of the muon's anomalous magnetic moment.  If measurements
of precession frequency are interpreted as measurements of $\amu$, the
assumption of a negligible muon EDM is best made explicit.
Alternatively, the experimental status may be summarized without
theoretical assumptions as in Fig.~\ref{fig:gm2exp}.


\end{document}